\documentclass[reprint, superscriptaddress, secnumarabic, amssymb, nobibnotes, aps, prl]{revtex4-1}

\usepackage{graphicx}
\usepackage{epstopdf}
\usepackage[T1]{fontenc}
\usepackage[latin9]{inputenc}
\usepackage{amsbsy}
\usepackage{gensymb}
\usepackage[T1]{fontenc}
\usepackage[latin9]{inputenc}
\usepackage{amsmath}
\usepackage{amssymb}
\usepackage{bbm}
\usepackage{physics}
\usepackage{braket}
\usepackage{xcolor}
\allowdisplaybreaks
\usepackage{graphicx}
\usepackage[colorlinks=true]{hyperref}  
\hypersetup{
    bookmarks=true,         
    unicode=false,          
    pdftoolbar=true,        
    pdfmenubar=true,        
    pdffitwindow=false,     
    pdfstartview={FitH},    
    pdftitle={Y3Ru4Ge13},    
    pdfauthor={},     
    pdfsubject={},   
    pdfcreator={},   
    pdfproducer={}, 
    pdfkeywords={} {} {}, 
    pdfnewwindow=true,      
    colorlinks=true,       
    linkcolor=blue, 
    citecolor=blue,        
    filecolor=magenta,      
    urlcolor=blue           
} 
\usepackage[normalem]{ulem}

\newcommand{\equref}[1]{Eq.~(\ref{#1})}

\newcommand{\figref}[1]{Fig.~\ref{#1}}
\newcommand{\refcite}[1]{Ref.~\onlinecite{#1}}

\renewcommand{\approx}{\simeq}

\begin{document}

\title{\textrm{Broken time-reversal symmetry in  cubic skutterudite-like superconductor Y$_3$Ru$_4$Ge$_{13}$}}
\author{A.~Kataria}
\affiliation{Department of Physics, Indian Institute of Science Education and Research Bhopal, Bhopal, 462066, India}
\author{J.~A.~T.~Verezhak}
\affiliation{Department of Physics, University of Warwick, Coventry, CV4 7AL, UK}
\author{O.~Prakash}
\affiliation{Department of Condensed Matter Physics and Materials Science,
Tata Institute of Fundamental Research, Mumbai-400005, India}
\author{R.~K.~Kushwaha}
\affiliation{Department of Physics, Indian Institute of Science Education and Research Bhopal, Bhopal, 462066, India}
\author{A.~Thamizhavel}
\affiliation{Department of Condensed Matter Physics and Materials Science,
Tata Institute of Fundamental Research, Mumbai-400005, India}
\author{S.~Ramakrishnan}
\affiliation{Department of Physics,  
Indian Institute of  Science Education and Research, Pune-411008, India}
\author{M.~S.~Scheurer}
\affiliation{Institute for Theoretical Physics III, University of Stuttgart, 70550 Stuttgart, Germany}
\author{A.~D.~Hillier}
\affiliation{ISIS Facility, STFC Rutherford Appleton Laboratory, Didcot OX11 0QX, United Kingdom}
\author{R.~P.~Singh}
\email[]{rpsingh@iiserb.ac.in}
\affiliation{Department of Physics, Indian Institute of Science Education and Research Bhopal, Bhopal, 462066, India}

\begin{abstract}
The microscopic properties of superconducting cubic skutterudite-like material Y$_3$Ru$_4$Ge$_{13}$ are investigated using muon spin relaxation and rotation ($\mu$SR) measurements. Zero-field $\mu$SR measurements reveal the presence of a spontaneous internal field with a magnitude of $\approx$ 0.18~mT below the superconducting transition temperature, indicating broken time-reversal symmetry in the ground state. In line with previous experiments, transverse-field $\mu$SR measurements are consistent with a fully developed superconductivity gap in Y$_3$Ru$_4$Ge$_{13}$. Our observations point towards the relevance of electronic correlations beyond electron-phonon coupling as origin and indicate that spin-orbit coupling is likely not the key driving force behind spontaneous breaking of time-reversal symmetry in this system.
\end{abstract}
\maketitle

\section{Introduction}

Breaking additional symmetries, such as time-reversal or rotational symmetry, in a superconductor is a defining feature of unconventional superconductivity and lies outside the scope of BCS theory \cite{bcs,bcs1}. In the superconducting state, the breaking of the time-reversal symmetry (TRS) is manifested by the spontaneous presence of a magnetic field around inhomogeneities below the superconducting transition temperature. This additional symmetry breaking in superconductivity can contribute to novel properties and rich and exciting physics, which motivates a detailed investigation of the ground state and attracts tremendous research attention. Studying such broken symmetries in superconductors is also crucial for understanding the pairing mechanism and the structure of the order parameter, as the symmetry of the system highly influences the order parameter symmetry and can lead to a nontrivial gap structure. 
\begin{figure*}[ht!]
\centering 
\includegraphics[width=2.0\columnwidth, origin=b]{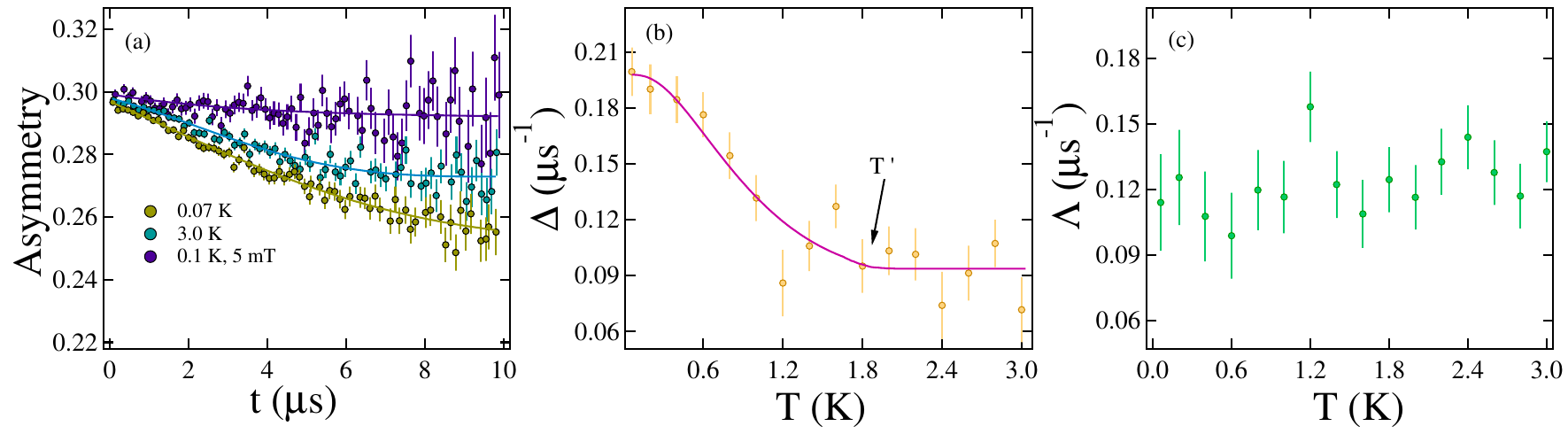}
\caption{\label{Fig1:ZF} (a) ZF asymmetry spectra at 0.07 K ($T<T_c$) and 3.0 K ($T>T_c$) with LF asymmetry spectra at 0.1 K and 5 mT, where solid lines represent the respective fit. (b) The relaxation rate $\Delta$, see \equref{eq1:kt}, increases at a temperature $T'$ below the superconducting $T_c$ (see the text for details), while the rate $\Lambda$ in \equref{eq2:zf} stays approximately constant (c).}
\end{figure*}

The phenomenon of broken TRS has been extensively investigated in cage-type superconducting systems due to their intriguing properties and complex structure. In this family, PrOs$_4$Sb$_{12}$ is reported as the first strongly correlated heavy-fermion superconductor with broken TRS, a complex order parameter, and multi-gap nodal superconductivity \cite{po4s12, po4s12_2}. Further studies of sister compounds with doping suggest the importance of Pr ion excitations, Pr-Pr interactions, and spin-orbit coupling (SOC) for TRS breaking in the superconducting state \cite{pp4g12,pro4s12,pcp4g12,plaos4sb12,pro4s12,preffect,prexcitions,plaos4sb12,lp4g12,tp4g12,lr4a12}. In contrast to this, PrRu$_4$Sb$_{12}$ exhibits no signs of broken TRS \cite{plaos4sb12,pr4s12}. Similarly, other cage compounds PrA$_2$M$_{20}$, (where A = V, Ir, Rh and M = Al, Zn), do not show evidence of broken TRS, further complicating the understanding of TRS breaking in cage compounds \cite{prm2x20,prm2x20_2,prm2x20_muon}. Meanwhile, tetragonal cage compounds of the form R$_5$Rh$_6$Sn$_{18}$ (R = Sc, Y, Lu) exhibit unconventional superconducting characteristics, including broken TRS, anisotropic gap structure, and rattling vibrations \cite{s5rh6s18,l5rh6s18,l5rh6s18_2,y5rh6s18}. Yet again, the ruthenate composition, R$_5$Ru$_6$Sn$_{18}$ shows BCS superconductivity and preserved TRS \cite{s5ru6s18}. Given this complex behavior and the limited number of studies on the cage structure family revealing different results, there is a huge gap that has to be overcome to be able to understand the superconducting order parameter, the mechanisms behind TRS breaking, and the pairing mechanism. The role of crystal structure, electron correlation, SOC, and rattling modes on the unconventional properties of cage-type superconductors remains largely unexplored, motivating a thorough examination of ground-state properties. 

In this context, the cubic skutterudite-like family R$_3$M$_4$S$_{13}$, where R represents a rare earth metal, M is a transition metal and S is a group-14 element, provides an exciting platform for investigating superconductivity in skutterudite structures \cite{rp}. Due to the strong covalent bonding in the cage R$_3$M$_4$S$_{13}$ structure, the rattling modes of small atoms are absent \cite{strongcovalent}. The weakly correlated stannides within this family exhibit unconventional properties with a BCS coupling mechanism, which provides another intriguing aspect of the study \cite{stanmuon,stanmuon2}. The strong interaction of superconductivity and crystal structure in R$_3$M$_4$Sn$_{13}$ is also suggested by the proximity of structural transition and superconductivity \cite{stan1,stan2}. Furthermore, recent investigations on the low-carrier-density superconductor Lu$_3$Os$_4$Ge$_{13}$ have revealed a multi-gap superconducting state with broken TRS, indicating the possibility of electron-electron interaction as a pairing mechanism \cite{y3r4g13_tdo,l3o4g13}. Therefore, the sister compound Y$_3$Ru$_4$Ge$_{13}$, with low-carrier-density, low SOC, and few or no rattling modes, offers a promising avenue to understand the effect of carrier concentration and SOC on the pairing mechanism of the cubic skutterudite family \cite{old_y3r4g13, y3r4g13,y3r4g13_2,y3r4g13_3}. 
Taken together, this motivates a microscopic investigation of the superconducting state of Y$_3$Ru$_4$Ge$_{13}$ using tools such as muon spin rotation/relaxation ($\mu$SR) measurements to present this compound in the context of other cage-type superconductors. Furthermore, understanding the possible pairing mechanisms, order parameters, and impact of various parameters, such as SOC, on TRS breaking would be crucial for cage compounds. To help fill this gap, this paper investigates the microscopic superconducting properties of the low-carrier-density system Y$_3$Ru$_4$Ge$_{13}$ via $\mu$SR spectroscopy. Zero-field (ZF) $\mu$SR measurements suggest broken TRS, as indicated by an increased relaxation rate below the superconducting transition temperature. As the associated magnetic field scale is even slightly larger than in the sister compound Lu$_3$Os$_4$Ge$_{13}$ \cite{l3o4g13} with stronger SOC, we conclude that SOC is not essential for stabilizing a TRS-breaking superconducting order parameter in these systems. 


\section{Experimental Details}

The single-crystals of Y$_3$Ru$_4$Ge$_{13}$ used for $\mu$SR measurements have already been characterized and studied in previous work \cite{y3r4g13,y3r4g13_2,y3r4g13_3}. Phase purity and superconducting properties were investigated using resistivity, magnetization, and specific heat measurements. $\mu$SR measurements have been performed on single crystals oriented along the [110] axis, with the incident muon beam always parallel to the [110] axis at the ISIS Neutron and Muon Pulsed Source, Appleton Laboratory, U. K. The methodology and instruments are detailed in \refcite{muon,muon2}. In the longitudinal configuration, measurements were performed with both an applied field (LF) and zero field (ZF), to infer the presence of an internal field in the superconducting state. While in the transverse field (TF) measurements, a magnetic field was applied perpendicular to the incident muon beam direction to investigate the gap structure in the vortex state of a superconductor.

\begin{figure*}
\centering
\includegraphics[width=2.0\columnwidth, origin=b]{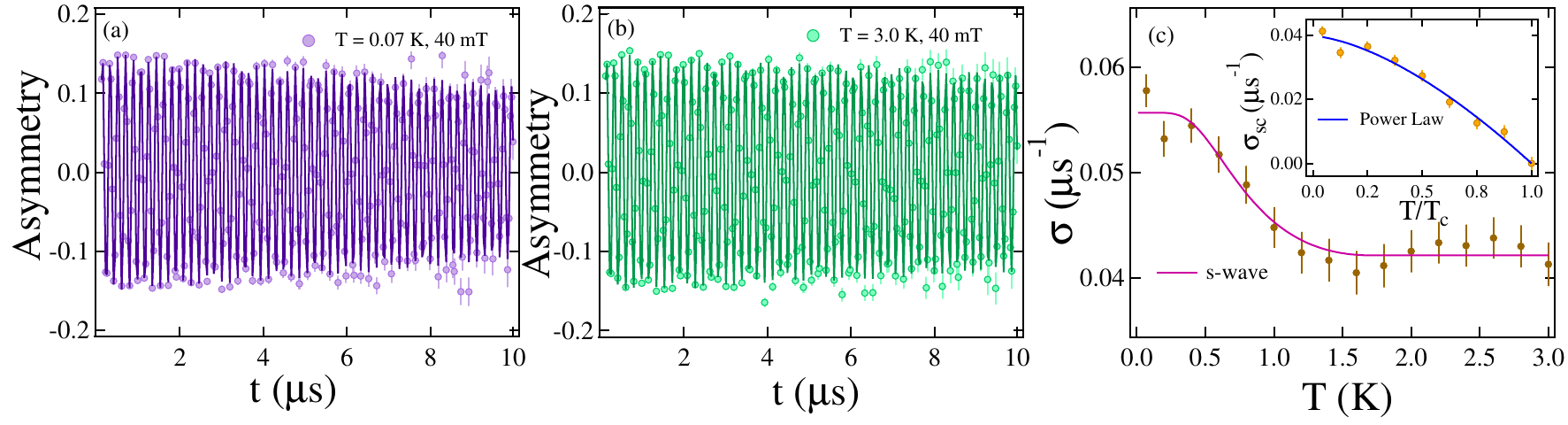}
\caption{\label{Fig2:TF} TF-$\mu$SR asymmetry spectra at 40 mT (a) in the superconducting state (at $T= 0.07\,\text{K}$) and (b) in the normal state, $T=3.0\,\text{K}$, with the solid lines representing the fitting using \equref{eqn3:TF}. (c) The muon relaxation rate, $\sigma$ variation with temperature fitted with the $s$-wave model with inset showing the fitting of $\sigma_{sc}$ with temperature using the power law. } 
\end{figure*}

\section{Result}

\subsection {Zero-Field $\mu$SR}
In ZF-$\mu$SR measurement, the time-domain asymmetry spectra were measured in regular intervals from above the superconducting transition temperature to the lowest temperature. The measured ZF asymmetry spectra above and below $T_c$ are shown in \figref{Fig1:ZF}(a). The observed small relaxation and absence of muon precession signal rule out the possibility of any long-range magnetically ordered state. The asymmetry spectra are analyzed by considering relaxation due to static, randomly oriented nuclear moments and fast fluctuation due to electronic moments. This can be expressed by the exponentially damped Gaussian Kubo-Toyabe (KT) function with a flat background \cite{zff},
\begin{equation}
     A(t)= A_0 G_z(t)\exp(-\Lambda t)+ A_{bg}
     \label{eq2:zf}
\end{equation}
where 
\begin {equation}
G_z(t)= \frac{1}{3} + \frac{2}{3}(1-\Delta^2t^2)\exp\left(-\frac{\Delta^2t^2}{2}\right)
\label{eq1:kt}
\end {equation} 
is the Gaussian Kubo-Toyabe function with $\Delta$ encoding the relaxation due to static or quasistatic local fields \cite{kt}; $A_0$ and $A_{bg}$ are the initial and background asymmetries, respectively, and $\Lambda$ represents the electronic relaxation rate. While $A_0$ and $A_{bg}$ are temperature independent, \figref{Fig1:ZF}(b) shows the increase in $\Delta$ with temperature around the superconducting transition temperature corresponding to an increase in the typical local magnetic fields \cite{old_y3r4g13}. Further, an applied longitudinal field of 5 mT was sufficient to decouple the muons from the decaying channel [shown in \figref{Fig1:ZF}(a)], indicative of the static or quasi-static nature of the field and broken TRS in the superconducting state of Y$_3$Ru$_4$Ge$_{13}$. A similar increase in $\Delta$ and broken TRS is also observed in PrOs$_4$Sb$_{12}$ \cite{po4s12}, PrPt$_4$Ge$_{12}$ \cite{pp4g12}, LaNiGa$_2$ \cite{lng2} and many Re-based superconductors \cite{r6z,r6h,r6ti,r6nb,re2hf}. The temperature dependence of $\Delta$ [\figref{Fig1:ZF}(b)] suggests the Gaussian-type internal field distribution in the superconducting state, which is different from other known TRS-breaking superconductors such as Sr$_2$RuO$_4$ \cite{s2ro4}. 

The measured change in the relaxation rate, $\Delta$ below $T_c$, is 0.11 $\mu$s$^{-1}$. The magnitude of the internal field associated with this change can be estimated by using the expression \cite{lmf},
\begin{equation}
    B_{int} = \sqrt{2}\frac{\Delta}{\gamma_{\mu}}
\end{equation}
where $\gamma_{\mu}$ = 2$\pi$ $\times$ 135.5 MHz/T is the muon gyromagnetic ratio. The estimated value of $B_{int} \approx$ 0.18(3) mT is comparable to reported values of 0.12~mT for filled skutterudite PrOs$_4$Sb$_{12}$ \cite{po4s12}, 0.116~mT for Re$_2$Hf \cite{re2hf} and 0.11~mT for Lu$_3$Os$_{4}$Ge$_{13}$ \cite{l3o4g13}. However, this value is considerably larger than that of other known TRS-breaking superconductors.

Furthermore, in \figref{Fig1:ZF}(b), it is observed that the actual increase in $\Delta$ occurs at a temperature of $T'$ = 1.7~K, which is below the resistive superconducting transition temperature of $T_c = 2.8$ K with a transition width of 0.4 K. Measurement of AC susceptibility of \refcite{y3r4g13_tdo} indicated the onset of the superconducting transition at a temperature of 2.2 K in the same sample of Y$_3$Ru$_4$Ge$_{13}$. Additionally, previous studies suggest a transition temperature ranging from 1.4 K to 1.7 K for this sample \cite{old_y3r4g13}. The discrepancy in the observed $T_c$ values may be attributed to the broad superconducting transition, the point considered as the transition temperature, and also the sensitivity of the measurement technique.


\subsection {Transverse-Field $\mu$SR}

The superconducting gap structure of Y$_3$Ru$_4$Ge$_{13}$ can be evaluated from the TF-$\mu$SR measurements. Asymmetry spectra were measured in a field-cooled protocol up to a temperature above $T_c$ at fixed intervals. \figref{Fig2:TF}(a) and (b) show the TF asymmetry spectra above and below $T_c$ under an applied magnetic field of 40 mT. No significant changes can be found in the two asymmetry spectra. The time-domain TF asymmetry spectra were further analysed by fitting the Gaussian-damped oscillatory relaxation function \cite{tff,tff1},
\begin{multline}
A (t) = A_{1}\exp\left(-\frac{1}{2}\sigma^2t^2\right)\cos(\gamma_\mu B_1t+\phi) \\ + A_2 \cos(\gamma_\mu B_{2}t +\phi)
\label{eqn3:TF}
\end{multline}
where $A_1$ and $A_2$ are the initial asymmetries of the sample and background, respectively. $\sigma$ is the Gaussian relaxation rate, $\phi$ is the initial phase, and $B_1$ and $B_2$ are the local magnetic field sensed by the muons in the sample and in the sample holder (background), respectively. The solid line represents the fitting of the spectra in \figref{Fig2:TF}(a) and (b). The extracted variation of $\sigma$ with temperature is shown in \figref{Fig2:TF}(c), revealing only a small change in $\sigma$ below $T_c$ within the experimental resolution of the time window $\mu$SR. The temperature dependence of $\sigma$ has been modelled (solid red line) using a single nodeless $s$-wave model in the clean limit \cite{s-wave}, providing a superconducting energy gap, $\Delta(0)$ = 0.17(1) meV. The obtained value differs from that reported in other measurements, including the specific heat (0.21 meV) \cite{y3r4g13} and tunnel-diode oscillator measurements ( $\approx$ 0.40 meV) \cite{y3r4g13_tdo}.

The small change in $\sigma$ versus $T$ below $T_c$ suggests that the vortex lattice has only a minor relaxation contribution, $\sigma_{sc}$. An upper bound can be estimated using $\sigma_{sc}$ = ($\sigma^2$- $\sigma_{dip}^2$)$^{1/2}$, where $\sigma_{dip}$ is the relaxation due to the nuclear moments measured in the normal state. The weak temperature dependence of the nuclear relaxation rate is also observed in La$_2$(Cu$_{1-x}$Ni$_x$)$_5$As$_3$O$_2$ similar to our case \cite{sn_dip}. We fit the temperature variation of $\sigma_{sc}$ with $T_c$ = 1.6 K using the power law (see \cite{2fm,2fm2}), 
\begin {equation}
\sigma_{sc}= \sigma_{sc}(0) \left[1-\left(\frac{T}{T_c}\right)^N
\right] \label{PhenModel}
\end {equation}
for $T \leq T_c$. The fitting yields $N = 1.56(5)$, much less than the value 4 expected for BCS superconductors. In dirty $d$-wave superconductors systems, it is suggested that $N \sim 2$ \cite{sn_dip, dn1, dn2}. The value for $N$ we obtain points towards a nodal superconducting order parameter, which is surprising and in contrast with the microscopic model. Thus, these small temperature variations of $\sigma$ and low superconducting gap values with relatively large uncertainties are inadequate to accurately determine the superconducting gap symmetry. We leave this for future work. 

Furthermore, considering the vortex lattice system of a type-II superconductor, an estimation of London penetration depth, $\lambda (0)$, can be extracted by using the expression \cite{lamvor},
\begin {equation}
\frac{\sigma_{sc} (T)}{\gamma_{\mu}} = 0.0609 \frac{\Phi_0}{\lambda^2(T)}.
\label{eqn4:si}
\end {equation}
The calculated lower bound is $\lambda (0)$ $\geq$ 1.6 $\mu$m. 
Moreover, using London's equation, the lower limit of the superconducting carrier density, $n$, can be extracted via the relation $\lambda^2= m^*/ \mu_0 n e^2$, where $m^*$ is the effective mass (taken from \refcite{y3r4g13}); the resulting bound $n$ $\leq$ 1.7 $\times$ 10$^{25}$ carriers/m$^{3}$ justifies the low-carrier density limit of the system. 

\begin{figure}
\includegraphics[width=0.95\columnwidth]{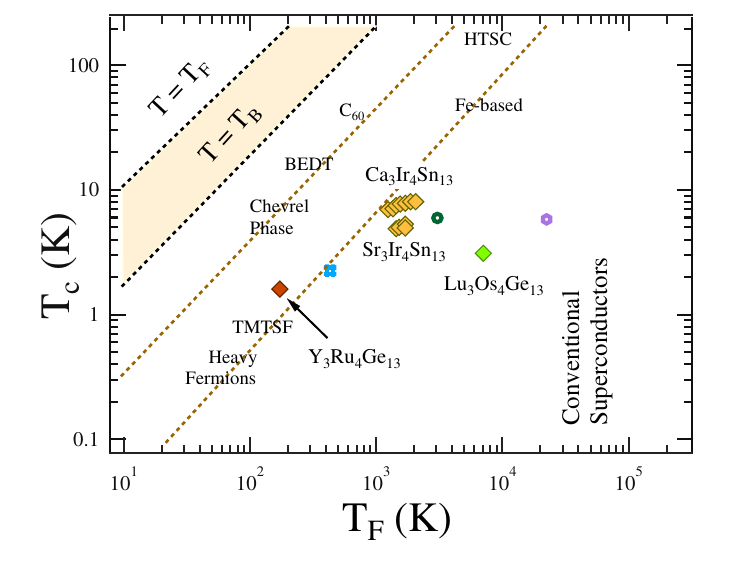}
\caption{\label{Fig3} Uemura plot of unconventional superconductors with Y$_3$Ru$_4$Ge$_{13}$ and other stannide of the family Ca$_3$Ir$_4$Sn$_{13}$ and Sn$_3$Ir$_4$Sn$_{13}$ at different pressure \cite{stanmuon, Unconv_1,Unconv_2}.} 
\end{figure}

According to Uemura's classification system, which is based on the ratio of the Fermi temperature to the superconducting transition temperature, $T_{\mathrm{F}}/T_c$, Y$_3$Ru$_4$Ge$_{13}$ is categorized with other unconventional superconductors. To obtain an estimate, we assume a 3D spherical Fermi surface and evaluate the Fermi temperature $T_F$ from \cite{tf},
\begin{equation}
k_{B}T_{\mathrm{F}} = \frac{\hbar^{2}}{2}(3\pi^{2})^{2/3}\frac{n^{2/3}}{m^{*}}, 
\label{eqn16:Tf}
\end{equation}
where $n$ is the carrier density, and $m^{*}$ is the effective mass and resultant $T_{\mathrm{F}}$ = 173(25) K. Figure \ref{Fig3} shows that Y$_3$Ru$_4$Ge$_{13}$ is positioned well inside the unconventional superconductor band.

A small change in $\sigma$ below $T_c$ or no significant muon relaxation in the superconducting state has previously been observed in the Cu or Sr-doped topological insulator Bi$_2$Se$_3$ \cite{srb2s3,cub2s3}, Ru$_{0.75}$Rh$_{0.25}$As \cite{RRP}, noncentrosymmetric YPtBi \cite{ypb} and most recently in CaSn$_3$ \cite{cs3}. The common features among the materials mentioned above and Y$_3$Ru$_4$Ge$_{13}$ are low-carrier-density and nontrivial band topology. In the Heusler alloy YPtBi and A15 compound CaSn$_3$, nontrivial topological band structures have recently been recognized \cite{cs3_ts}. Moreover, theoretical band structure calculations from the topological material database also suggest Y$_3$Ru$_4$Ge$_{13}$ as a symmetry-enforced semimetal \cite{tqc,tqc2,tqc3}. Additionally, peak effect and anomalous susceptibility behavior with the magnetic field are observed for Y$_3$Ru$_4$Ge$_{13}$  \cite{y3r4g13_tdo}. These observations suggest weak vortex pinning at a low magnetic field and hint toward the possibility of a complex vortex lattice and field-induced disordering of the vortex lattice \cite{y3r4g13_tdo}. However, the precise mechanism governing the unconventional vortex state in low-carrier topological materials remains elusive. Further investigation in this direction is imperative to understand the intricacies of the vortex-state phenomenon.

\section{Discussion and Conclusion}
Finally, we discuss the implications of our findings for the microscopic pairing state and mechanism in Y$_3$Ru$_4$Ge$_{13}$. While it has the same symmetries as Lu$_3$Os$_4$Ge$_{13}$---both exhibit the space group no.~223 ($Pm\bar{3}n$) with the cubic point group $O_h$---the expected strength of SOC ($\propto Z^4$) is much weaker in Y$_3$Ru$_4$Ge$_{13}$, given the significantly lower atomic numbers of the involved atoms; more precisely, first principle calculations indicate that the states around the Fermi level in Y$_3$Ru$_4$Ge$_{13}$ derive mostly from Ge ($Z=32$) \cite{FPYRuGe} and those in Lu$_3$Os$_4$Ge$_{13}$ also exhibit significant contributions from Os ($Z=76$) \cite{FPLuOsGe}. Yet, the key observation of TRS breaking setting in at a temperature below the resistive superconducting transition is common in the present and our previous \cite{l3o4g13} $\mu$SR study of Y$_3$Ru$_4$Ge$_{13}$ and Lu$_3$Os$_4$Ge$_{13}$, respectively. In combination with the similar strength of the internal field $B_{int}$ in these compounds, this strongly indicates that SOC is not the key driving force behind the TRS breaking.  

By construction, the symmetry-based discussion of the possible pairing states for Lu$_3$Os$_4$Ge$_{13}$ presented in \refcite{l3o4g13} applies here as well: if the superconducting state is reached by a single phase transition, the order parameter must transform under a two- or three-dimensional irreducible representation of $O_h$ to allow for broken TRS. As a result of the complex structure of the group $O_h$, there are ten distinct such TRS-breaking superconducting candidate phases; generically, all of them are expected to exhibit nodes. As pointed out above, our TF $\mu$SR data are consistent with a full gap, but the small variation of $\sigma$ might not be enough to rule out the presence of nodes. 
Since, however, more sensitive tunnel-diode oscillator measurements also indicate a fully gapped superconducting phase \cite{y3r4g13_tdo}, alternative scenarios leading to a superconducting order parameter with broken TRS and a full gap should be discussed. Given that the signs of broken TRS in both Y$_3$Ru$_4$Ge$_{13}$ and Lu$_3$Os$_4$Ge$_{13}$ emerge at a temperature $T^\prime$ that is noticeably smaller than the resistive $T_c$, the most plausible scenario is that there are two superconducting transitions---at the first one, a (nodal or fully gapped) state sets in while only at a lower temperature, $T^\prime<T_c$, a secondary order parameter appears, which, e.g., due to a non-trivial relative complex phase, breaks TRS; at least below $T^\prime$, the superconducting order parameter has a full gap. Such a TRS-breaking complex phase can arise due to ``frustrated'' interactions (see Refs.~\cite{l7n3,ChubukovComplSC,ChubukovComplSCEliashberg} for example). Irrespective of which of these scenarios is realized, we expect electronic interactions beyond electron-phonon coupling to be present in order to stabilize the TRS-breaking superconducting phase below $T^\prime$, making Lu$_3$Os$_4$Ge$_{13}$ a promising candidate for unconventional superconductivity.

In summary, the microscopic properties of the superconducting cubic skutterudite-like material Y$_3$Ru$_4$Ge$_{13}$ were investigated using muon spin rotation and relaxation measurements. Zero-field $\mu$SR results suggest TRS breaking in the superconducting ground state, similar to its Lu sister compound and many other skutterudite compounds. A significant local internal magnetic field of magnitude $\approx$ 0.18(3) mT is observed in the superconducting state. TF-$\mu$SR measurements are consistent with a fully gapped superconducting order parameter. Our findings indicate that SOC is likely not the key driving force for TRS-breaking superconductivity and that pairing interactions beyond the conventional electron-phonon coupling are at play. Nonetheless, there are still many open questions that deserve further investigations, such as determining the form of the superconducting order parameter and understanding the nature of the vortex state in Y$_3$Ru$_4$Ge$_{13}$. More generally, our work emphasizes that low-carrier-concentration systems and cage compounds are an intriguing playground for exotic superconducting physics.  

\section{Acknowledgments}
A. Kataria acknowledges the CSIR, Government of India, for the SRF fellowship (Award No: 09/1020(0172)/2019-EMR-I). R.~P.~S. acknowledges the SERB, Government of India, for the Core Research Grant CRG/2019/001028, and ISIS, STFC, UK, for providing beamtime for the $\mu$SR experiments.

\end{document}